\documentclass[10pt,conference]{IEEEtran}

\usepackage{balance}
\usepackage{algorithm}
\usepackage{array}
\usepackage[caption=false,font=normalsize,labelfont=sf,textfont=sf]{subfig}
\usepackage{algpseudocode}
\usepackage{textcomp}
\usepackage{csquotes}
\usepackage{tabstackengine}
\usepackage{multirow}  
\usepackage{booktabs}  
\usepackage{threeparttable}  
\usepackage{url}
\usepackage{verbatim}
\usepackage{graphicx}
\usepackage{cite}
\usepackage{amssymb,amsmath,amsthm,enumitem}

\usepackage[cmintegrals]{newtxmath}
\usepackage{siunitx}
\usepackage[nolist]{acronym}
\usepackage{tikz}
\usetikzlibrary{shapes.geometric, arrows.meta, positioning, calc}

\usepackage[export]{adjustbox} 
\usepackage[left=1.62cm,right=1.62cm,top=1.9cm]{geometry}
\begin{document}

\title{EVT-Based Generative AI for Tail-Aware Channel Estimation}
\author{
    \IEEEauthorblockN{
    Parmida Valiahdi\IEEEauthorrefmark{1}, 
    Niloofar Mehrnia\IEEEauthorrefmark{2},
    Walid Saad\IEEEauthorrefmark{3},
    Sinem Coleri\IEEEauthorrefmark{1}}
    
    \IEEEauthorblockA{\IEEEauthorrefmark{1}Department of Electrical and Electronics Engineering, Koc University, Istanbul, Turkey}
    \IEEEauthorblockA{\IEEEauthorrefmark{2}Department of Electrical Engineering and Computer Science, KTH Royal Institute of Technology, Stockholm, Sweden}

    \IEEEauthorblockA{\IEEEauthorrefmark{3}Department of Electrical and Computer Engineering, Virginia Tech, Virginia, United States}
}




\maketitle
\begin{acronym}
    \acro{URLLC}{Ultra-Reliable Low Latency Communication}
  \acro{EVT}{Extreme Value Theory}
  \acro{GPD}{Generalized Pareto Distribution}
  \acro{RMSE}{Root Mean Square Error}
  \acro{MAE}{Mean Absolute Error}
  \acro{MLE}{Maximum Likelihood Estimation}
  \acro{GAN}{Generative Adversarial Network}
  \acro{EVT-HGAN}{EVT-guided Hybrid Generative Adversarial Network}
  \acro{VAE}{Variational Autoencoder}
  \acro{MLP}{Multilayer Perceptron}
  \acro{POT}{Peak-Over-Threshold}
  \acro{GEV}{Generalized Extreme Value}
  \acro{QQ}{Quantile-Quantile}
  \acro{KS}{Kolmogorov-Smirnov}
  \acro{PPCC}{Probability Plot Correlation Coefficient}
  \acro{CDF}{Cumulative Distribution Function}
  \acro{MSE}{Mean Squared Error}
  \acro{GRU}{Gated Recurrent Unit}
  \acro{LSTM}{Long Short-Term Memory}
  \acro{RNN}{Recurrent Neural Network}
  \acro{DRNN}{Deep Recurrent Neural Network}
  \acro{Tx}{Transmitter}
  \acro{Rx}{Receiver}
  \acro{AI}{Artificial Intelligence}
  \acro{DM}{Diffusion Models}
  \acro{KL}{Kullback-Leibler}
  \acro{BCE}{Binary Cross-Entropy}
  \acro{MIMO}{Multiple-Input Multiple-Output}
  \acro{CSI}{Channel State Information}
  \acro{GMM}{Gaussian Mixture Model}
  \acro{BIC}{Bayesian Information Criterion}
  \acro{SNR}{Signal-to-Noise Ratio}
  \acro{GPU}{Graphics Processing Unit}
\end{acronym}
\begin{abstract}
Ultra-reliable and low-latency communication (URLLC) will play a key role in fifth-generation (5G) and beyond networks, enabling mission-critical applications. Meeting the stringent URLLC requirements, characterized by extremely low packet error rates and minimal latency, calls for advanced statistical modeling to accurately capture rare events in wireless channels. Traditional methods, such as those that rely on large datasets and computationally intensive estimation techniques, often fail in real-time scenarios. In this paper, a novel framework is proposed to meet URLLC requirements through a synergistic integration of extreme value theory (EVT) with generative artificial intelligence (AI).  
EVT is used to model channel tail distributions, providing an accurate characterization of rare events. Concurrently, generative AI enables data augmentation and channel parameter estimation from limited samples. 
The integration of EVT with generative AI can thus help overcome the limitations of generative models in capturing extreme events during channel characterization.
Using an experimental dataset collected from an automotive environment, it is demonstrated that this integration enhances data augmentation for extreme quantiles, while requiring fewer samples than traditional analytical EVT methods and generative baselines in online estimation of channel distribution.
\end{abstract}


\section{Introduction}
\label{sec:introduction}
\ac{URLLC} is a fundamental technology that enables mission-critical applications, including remote robotic control, autonomous vehicles, and telemedicine, especially in 5G and future wireless networks~\cite{urllc_01}. \ac{URLLC} aims to achieve extremely low packet error rates, ranging from $10^{-9}$ to $10^{-5}$, while maintaining end-to-end latencies of just a few milliseconds or even less. This means that system performance is often determined by rare events where the received signal power is extremely low, leading to outages.
Capturing and accurately modeling these extreme events is a significant challenge. Simply gathering large datasets and fitting them to standard statistical distributions often fails to represent the tail behavior, due to the scarcity of such events and mismatches between models and real environments. As a result, there is a growing need for channel models that specifically focus on the tail of the distribution, as well as lightweight, online inference techniques. 

\ac{EVT} is a robust statistical framework that has been explicitly designed to model rare events and to provide characterizations of the distribution tail \cite{scoles}. In wireless communications, EVT has been used recently to quantify extreme interference, outage, and delay via two complementary approaches: the block maxima/minima method, typically modeled by the \ac{GEV} distribution, and the \ac{POT} method, which models exceedances using the \ac{GPD} \cite{saad_evt,evt_01,evt_02,evt_03}. A novel \ac{EVT}-based channel modeling methodology has been proposed in \cite{evt_01} to derive ultra-reliable channel statistics from limited data with high precision. The \ac{EVT} framework has been extended to non-stationary wireless channels in \cite{evt_02} and to multi-channel wireless systems in \cite{evt_03}. Additionally, \cite{saad_evt} applies \ac{EVT} using the block maxima (or minima) approach, where the data is divided into fixed-size blocks, and the extreme value from each block is modeled using the \ac{GEV} distribution. Recently, EVT has been incorporated into the loss function of deep recurrent neural networks (DRNN) in \cite{mehrnia2025evt} to predict extreme quantiles, where \ac{GPD} parameters are estimated offline and then fed into the loss function for the prediction. However, the practical accuracy of \ac{EVT} depends critically on how well one can select thresholds and estimate tail parameters in real time. These tasks become challenging in real time when only a small number of extreme samples are available. This gap motivates integrating \ac{EVT} with generative \ac{AI}, leveraging learned generative models to capture complex distributional structure and to support data-efficient, adaptive estimation of \ac{EVT} tail parameters under stringent \ac{URLLC} latency constraints.

Generative \ac{AI}, notably \acp{GAN} and \acp{DM}, has emerged as a powerful data-driven tool for learning complex wireless channel statistics, enabling channel modeling, denoising, and inference when analytical models are inaccurate or unavailable \cite{genai_05,genai_11,genai_04,diff_1,diff_2}. At the physical layer, these models are trained to match measured channel distributions: \ac{GAN}s learn through an adversarial generator, discriminator game, while \ac{DM}s learn to reverse a progressive noise-corruption process via iterative denoising, often providing strong sample quality and improved mode coverage \cite{diff_1}. Prior work has used offline-trained GANs as learned priors for channel estimation in low \ac{SNR} or pilot-limited regimes \cite{genai_05,genai_csi_01}, extended GAN-based designs to channel prediction by combining them with \ac{LSTM}s to capture temporal structure \cite{genai_04}, and for high-dimensional \ac{MIMO}/wideband \ac{CSI} recovery \cite{genai_gan_11,genai_gan_12}. More recently, diffusion-based methods have been advocated for channel generation, estimation, and denoising due to robustness under distribution shift \cite{diff_11, genai_fl_01}. Despite these advances, prevailing objectives remain focused on average reconstruction error (e.g., \ac{MSE}) rather than \ac{URLLC}-critical extreme-tail fidelity.

This paper proposes an \ac{EVT}-guided generative learning framework for real-time estimation of \ac{URLLC}-relevant channel tail behavior. The central idea is to combine the principled extreme-tail modeling of \ac{EVT} with the adaptability of generative AI, so that online tail inference does not drift toward average-case accuracy and does not require impractically large datasets to capture rare outages. This architecture extends the \ac{GAN}-based estimation approach for \ac{GPD} shape and scale parameters proposed in \cite{genai_evt_01} by incorporating optimal threshold selection and real-time GPD parameter estimation derived from \ac{EVT} principles, thereby achieving greater robustness in non-stationary environments. The main contributions of the paper are summarized as follows:

\begin{itemize}
    \item We embed \ac{EVT} tail modeling into a generative-\ac{AI}–based estimator so that the learned model explicitly targets rare-event statistics rather than only fitting the bulk of the distribution.
    \item We design a methodology to select an optimal threshold separating extreme from non-extreme samples, estimate tail-distribution parameters (i.e., \ac{GPD} parameters), and validate the tail fit using standard \ac{EVT} diagnostic checks, structured for repeated, real-time updates.
    \item The framework uses generative \ac{AI} to selectively enrich the extreme region, rather than uniformly augmenting data, and improve tail-parameter estimation accuracy when true tail observations are scarce. It reduces the number of real measurements required to obtain stable tail-parameter estimates suitable for \ac{URLLC} operation.
    \item We evaluate the performance of the proposed framework using the data obtained from an intra-vehicular communication at \(60\)~GHz and by comparing the framework with the state-of-the-art generative models.
\end{itemize}

The rest of the paper is organized as follows: Section~\ref{sec:system_model} provides the system model and assumptions used throughout the paper. In Section~\ref{sec:method}, we describe the proposed real-time estimation framework that integrates EVT with generative AI. Section~\ref{sec:numerical_results} provides the measurement setup and the numerical results. Finally, conclusions and future works are presented in Section~\ref{sec:conclusions}.

\section{System Model}
\label{sec:system_model}
We consider a wireless link whose quality is monitored through a scalar time series sampled every $T_s$ seconds. Let $x_t\in\mathbb{R}$ denote the measured received power (equivalently, squared channel gain) at discrete time index $t$. An outage (reliability violation) is declared when the received power falls below a service-dependent threshold,
\begin{equation}
\mathcal{O}_t \triangleq \{x_t \le x_{\mathrm{th}}\},
\end{equation}
where $x_{\mathrm{th}}$ is determined by the required link margin and the employed modulation/coding and receiver processing. Therefore, \ac{URLLC} performance is governed by the \emph{lower tail} of the received-power distribution.

Wireless measurements are generally temporally correlated. Since \ac{EVT}-based tail modeling is theoretically justified under (approximately) independent and identically distributed (i.i.d.) samples, we adopt a local-stationarity assumption: over short intervals, the distribution of \(x_t\) is approximately stationary. Within each interval, temporal dependence is mitigated using a preprocessing step (e.g., declustering or a lightweight whitening model), yielding an approximately i.i.d.\ sequence $\tilde{x}_t$. For notational convenience, we denote the preprocessed sample by
\(y_t \triangleq \tilde{x}_t\), and use $y_t$ as the input to the online estimator.

\subsection{Lower-Tail Modeling via \ac{POT}}
\label{subsec:pot}
To model rare \emph{low}-power events, we adopt a lower-tail peaks-over-threshold (\ac{POT}) formulation. Let $u$ denote a (low) tail threshold in the received-power domain, and define the \emph{deficit} (lower-tail exceedance)
\begin{equation}
Z \triangleq u - Y \;\;\; \text{conditioned on}\;\;\; Y<u,
\label{eq:deficit}
\end{equation}
where $Y$ denotes a generic random variable with the same distribution as $y_t$. Thus, $Z\ge 0$ measures how far a sample falls below the threshold.

Under the \ac{POT} theory, for sufficiently low $u$, the conditional distribution of the deficit is approximated by a \ac{GPD}:
\begin{equation}
\Pr(Z \le z \mid Y<u) \approx G(z;\xi,\beta)
= 1-\left(1+\frac{\xi z}{\beta}\right)^{-1/\xi},
\label{eq:gpd_cdf}
\end{equation}
where $z\ge 0$, and $\xi$ and $\beta>0$ are the shape and scale parameters, respectively, with support constraint $1+\xi z/\beta>0$.

\subsection{Online Tail-Parameter Estimation Objective}
\label{subsec:objective}
At each update time, the receiver observes a short, most-recent window of preprocessed samples and aims to produce real-time estimates \((\hat{u}_t,\hat{\xi}_t,\hat{\beta}_t)\), such that the induced \ac{GPD} tail model accurately represents the \ac{URLLC}-relevant lower tail under tight sample and latency constraints. Here $\hat{u}_t$ is the estimated lower-tail threshold and $(\hat{\xi}_t,\hat{\beta}_t)$ are the corresponding \ac{GPD} parameters. In the next section, we describe an \ac{EVT}-guided generative learning framework that jointly estimates $\hat{u}_t$ and $(\hat{\xi}_t,\hat{\beta}_t)$ from streaming windows, and trains the parameter estimates by explicitly matching the empirical and generated tail distributions.
\section{Proposed Online Tail-Aware Estimation}
\label{sec:method}

\begin{figure}[thb]
    \centering
    \begin{adjustbox}{width=\linewidth}

    \definecolor{csand}{HTML}{CBBBA8}
    \definecolor{cclay}{HTML}{BCA28D}
    \definecolor{csage}{HTML}{AAB8A4}
    \definecolor{cslate}{HTML}{8FA4A8}
    \definecolor{clsand}{HTML}{D7CABC}
    \definecolor{ctaupe}{HTML}{C7B1A1}
    \definecolor{cpsage}{HTML}{B9C6B3}
    \definecolor{clslate}{HTML}{9DB1B5}
    \definecolor{cedge}{HTML}{000000}

    \begin{tikzpicture}[
        >=Stealth,
        block/.style={rectangle, draw=cedge, thick, align=center,
            minimum height=1.2cm, font=\rmfamily\large, text=cedge,
            rounded corners=2pt},
        data_block/.style={block, fill=cpsage!50},
        model_block/.style={block, fill=clslate!50},
        op_block/.style={block, fill=clsand},
        loss_block/.style={block, fill=ctaupe},
        database/.style={cylinder, shape border rotate=90, aspect=0.25,
            draw=cedge, thick, align=center, minimum height=1.5cm,
            font=\rmfamily\large, text=cedge, fill=cpsage!70},
        circle_node/.style={circle, draw=cedge, thick, align=center,
            font=\rmfamily\large, text=cedge, fill=cpsage!30},
        title_box/.style={rectangle, draw=cedge, thick,
            font=\rmfamily\large\bfseries, text=cedge,
            fill=csand!70, inner sep=6pt},
        frame_box/.style={rectangle, draw=cslate, thick, rounded corners=4pt},
        arrow/.style={->, thick, draw=cedge!50},
        line/.style={-, thick, draw=cedge!50},
        dashed_arrow/.style={->, dashed, thick, draw=cslate!200},
        dashed_line/.style={-, dashed, thick, draw=cslate}
    ]

    \draw[frame_box] (-2, 5.2) rectangle (18, -3.8);
    \draw[frame_box] (-2, -4.2) rectangle (18, -13.0);

    \node[title_box, anchor=north west] at (-1.7, 4.9)
        {EVT-GAN for tail-aware data augmentation (offline)};

    \node[circle_node, text width=1.5cm] (noise) at (0, 0) {Random\\noise};
    \node[model_block, text width=3.6cm] (gen) at (4, 0) {EVT-based Hybrid\\Generator};
    \node[database, text width=3.5cm] (real_db) at (5, 2.5) {Real channel data\\(received power)};
    \node[model_block, text width=3cm] (disc1) at (10, 0) {Discriminator};
    \node[loss_block, minimum height=1cm, text width=1.2cm] (loss1) at (14, 0) {Loss};
    \node[database, text width=2.5cm] (aug_db) at (11, -2.5) {Augmented\\dataset};

    \draw[arrow] (noise) -- (gen);

    \draw[dashed_arrow] (real_db.south) -- (real_db.south |- gen.north);

    \draw[line] (real_db.east) -- ++(0.5, 0) coordinate (r_drop1);
    \draw[arrow] (r_drop1) |- ($(disc1.west) + (0, 0.3)$);
    \draw[arrow] (gen.east) |- ($(disc1.west) + (0, -0.3)$);

    \coordinate (gen_out) at ($(gen.east) + (0.4, 0)$);
    \draw[arrow] (r_drop1) |- ($(aug_db.west) + (0, -0.3)$) node[pos=0.92, above] {$+$};
    \draw[arrow] (gen.south) |- ($(aug_db.west) + (0,0.5)$);


    \draw[arrow] (disc1.east) -- (loss1.west);

    \draw[dashed_line] (loss1.north) -- ++(0, 0.8) coordinate (l1_fb);
    \draw[dashed_arrow] (l1_fb) -| (disc1.north);
    \draw[dashed_arrow] (l1_fb) -| ($(gen.north) + (-1.5, 0)$);

    \node[title_box, anchor=north west] at (-1.7, -4.5)
        {EVT-GAN for tail threshold and parameter estimation (online)};

    \node[data_block, text width=4.5cm] (buffer) at (1, -8.5)
        {Estimator buffer\\$\mathbf{b}_t = [y_{t-N_w+1}, \dots , y_t]$};
    \node[model_block, text width=3.2cm] (thresh_mlp) at (6, -7.5) {Threshold MLP};
    \node[model_block, text width=3.2cm] (param_mlp) at (6, -9.5) {Parameter MLP};

    \node[op_block, text width=3.4cm] (extract) at (10.5, -7.5)
        {Extract exceedances\\$z=\hat{u}_t-y,\; y<\hat{u}_t$};
    \node[op_block, text width=3.4cm] (gpd) at (10.5, -9.5) {GPD Sampling};

    \node[model_block, text width=2.5cm] (disc2) at (14.4, -8.5) {Discriminator};
    \node[loss_block, minimum height=1cm, text width=1cm] (loss2_top) at (16.8, -7) {Loss};
    \node[loss_block, minimum height=1cm, text width=1cm] (loss2_bot) at (16.8, -8.5) {Loss};

    \node[data_block, text width=3.5cm] (online_model) at (10.5, -11.8)
        {online tail model\\$(\hat{u}_t, \hat{\xi}_t, \hat{\beta}_t)$};

    \draw[dashed_arrow] (aug_db.south) -- ++(0, -2.5) -| (buffer.north);

    \draw[line] (buffer.east) -- ++(0.5, 0) coordinate (buf_split);
    \draw[arrow] (buf_split) |- (thresh_mlp.west);
    \draw[arrow] (buf_split) |- (param_mlp.west);

    \draw[line] (thresh_mlp.east) -- ++(0.6, 0) coordinate (t_drop);
    \draw[arrow] (t_drop) -- (extract.west);
    \draw[arrow] (t_drop) |- ($(online_model.west) + (0, 0.2)$);

    \draw[line] (param_mlp.east) -- ++(0.4, 0) coordinate (p_drop);
    \draw[arrow] (p_drop) -- (gpd.west);
    \draw[arrow] (p_drop) |- ($(online_model.west) + (0, -0.2)$);

    \draw[line] (extract.east) -- ++(0.4, 0) coordinate (ext_split);
    \draw[arrow] (ext_split) |- (loss2_top.west);
    \draw[arrow] (ext_split) |- ($(disc2.west) + (0, 0.3)$);

    \draw[arrow] (gpd.east) -- ++(0.4, 0) |- ($(disc2.west) + (0, -0.3)$);

    \draw[arrow] (disc2.east) -- (loss2_bot.west);

    \draw[dashed_line] (loss2_top.north) -- ++(0, 0.6) coordinate (lt_fb);
    \draw[dashed_arrow] (lt_fb) -| (thresh_mlp.north);

    \draw[dashed_line] (loss2_bot.south) -- ++(0, -1.4) coordinate (lb_fb);
    \draw[dashed_arrow] (lb_fb) -| (disc2.south);
    \draw[dashed_arrow] (lb_fb) -| (param_mlp.south);

    \end{tikzpicture}
    \end{adjustbox}
    \caption{Proposed EVT-based generative channel tail estimation framework.}
    \label{fig:flowchart}
\end{figure}
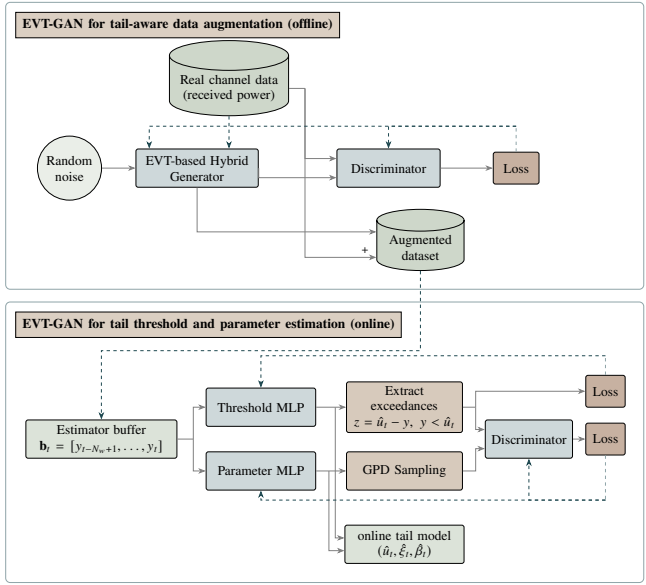

Given streaming received-power samples, the proposed framework produces an \emph{online} EVT tail model for \ac{URLLC} operation. As shown in Fig.~\ref{fig:flowchart}, the framework outputs at each update time $t$ the triplet
\begin{equation}
\widehat{\mathcal{M}}_t \triangleq \big(\hat{u}_t,\hat{\xi}_t,\hat{\beta}_t\big),
\end{equation}
where $\hat{u}_t$ and $(\hat{\xi}_t,\hat{\beta}_t)$ denote the estimated lower-tail threshold and the corresponding \ac{GPD} parameters, respectively. The framework comprises: (i) an \emph{offline} EVT-guided augmentation stage used only during training, and (ii) an \emph{online} tail-parameter estimation stage used during operation. 

\subsection{Offline EVT-Guided Tail Augmentation (Training Only)}
To handle nonstationarity, the stream is partitioned into regimes (batches) with approximately consistent tail behavior. Each window (defined below) is represented by a feature vector, and windows are clustered using a \ac{GMM} with the number of clusters selected via \ac{BIC} \cite{Fraley2002}. Windows assigned to the same cluster are treated as belonging to the same regime. During training, regime membership is used to pool tail samples and regularize window-level estimates.


To address the scarcity of extreme samples, we design an EVT-based hybrid generator that explicitly separates \emph{bulk} and \emph{tail} synthesis. Given a threshold $u$ and tail parameters $(\xi, \beta)$, synthetic samples are generated as
\begin{equation}
\tilde{y_t} =
\begin{cases}
u - \tilde{z}, & y_t < u, \\
g_{\phi}(\epsilon), & y_t \geq u,
\end{cases}
\end{equation}
where $\tilde{z} \sim \text{GPD}(\xi, \beta)$ and $g_{\phi}(\epsilon)$ denotes a neural generator driven by latent input $\epsilon$. Specifically, samples below the threshold are generated via parametric \ac{EVT}-based sampling to accurately capture extreme behavior, while samples above the threshold are generated by the neural component to model the bulk distribution. The resulting hybrid samples are then used within a standard \ac{GAN} framework, where the discriminator is trained to distinguish real channel samples from hybrid-generated samples (top panel of Fig.~\ref{fig:flowchart}).

\subsection{Online Tail Parameter Estimation}
\label{subsec:online_estimation}

Let $\{y_t\}_{t\ge 1}$ denote the streaming preprocessed received-power samples from Sec.~\ref{sec:system_model}. At each update time $t$, the receiver maintains an \emph{estimator buffer} of the most recent $N_w$ samples,
\begin{equation}
\mathbf{b}_t \triangleq [y_{t-N_w+1},\,y_{t-N_w+2},\,\ldots,\,y_t] \in \mathbb{R}^{N_w}.
\label{eq:buffer}
\end{equation}
We refer to $\mathbf{b}_t$ as the \emph{current window}. Windows are updated in a sliding manner as new samples arrive, hence consecutive windows typically overlap. A \emph{regime/batch} groups multiple windows with similar statistics.

Given a threshold $u$ and a set of samples $\mathcal{S}$, we form lower-tail exceedances as
\begin{equation}
z = u - y,\quad \text{for } y<u,\; y\in\mathcal{S}.
\label{eq:lower_tail_exceedance}
\end{equation}
When measurements are time-correlated, exceedances may appear in bursts. Therefore, a declustering operator \cite{scoles,evt_01} is applied to obtain approximately i.i.d.\ exceedances, denoted by $\mathcal{Z}$, which is then used for tail fitting and adversarial matching.

\subsubsection{Threshold Network}
Threshold is estimated from the current window using a threshold \ac{MLP} \cite{mlp}, \(\hat{u}_t = f_{\theta}^{(u)}(\mathbf{b}_t)\).
Using $\hat{u}_t$, exceedances are extracted from the window via \eqref{eq:lower_tail_exceedance} with $\mathcal{S}=\{y_{t-N_w+1},\ldots,y_t\}$, producing $\mathcal{Z}_t$.

The threshold network is trained using a tail-fit term and a tail-size penalty:
\begin{equation}
\mathcal{L}_{\mathrm{thr}}
=
\mathbb{E}_t\!\left[\mathcal{L}^{(t)}_{\mathrm{fit}}+\lambda_{\mathrm{tail}}\mathcal{L}^{(t)}_{\mathrm{tail}}\right].
\label{eq:thr_total}
\end{equation}
Here, $\mathcal{L}^{(t)}_{\mathrm{fit}}$ scores how well a \ac{GPD} fits the exceedances induced by $\hat{u}_t$. Specifically, we obtain provisional parameters $(\tilde{\xi}_t,\tilde{\beta}_t)$ from a lightweight per-window fit on $\mathcal{Z}_t$, used only to evaluate the threshold estimation, and define
\begin{equation}
\mathcal{L}^{(t)}_{\mathrm{fit}}
\triangleq
-\sum_{z\in\mathcal{Z}_t}\log p_{\mathrm{GPD}}(z;\tilde{\xi}_t,\tilde{\beta}_t),
\label{eq:thr_fit}
\end{equation}
where $p_{\mathrm{GPD}}(\cdot;\xi,\beta)$ is the \ac{GPD} density \eqref{eq:gpd_cdf}. The tail-size penalty discourages thresholds that yield too few exceedances:
\begin{equation}
\mathcal{L}^{(t)}_{\mathrm{tail}}
\triangleq
\big[\max(0,n_{\min}-|\mathcal{Z}_t|)\big]^2,
\label{eq:thr_tailpen}
\end{equation}
where $n_{\min}$ is the minimum desired number of exceedances for stable training.

\subsubsection{Parameter Network and Adversarial Tail Matching}
The tail parameters are estimated from the same window using a parameter MLP, \((\hat{\xi}_t,\hat{\beta}_t) = f^{(p)}_{\theta}(\mathbf{b}_t)\), with $\hat{\beta}_t>0$ enforced by the output activation, and $\hat{\xi}_t$ optionally bounded to a numerically stable range.

In the online EVT-GAN block (bottom panel of Fig.~\ref{fig:flowchart}), the generator is the composition
\begin{equation}
\mathbf{b}_t
\;\xrightarrow{\;f_{\theta}^{(p)}\;}\;
(\hat{\xi}_t,\hat{\beta}_t)
\;\xrightarrow{\;\mathrm{GPD\ sampling}\;}\;
\tilde{\mathcal{Z}}_t,
\label{eq:generator_comp}
\end{equation}
where $\tilde{\mathcal{Z}}_t=\{\tilde{z}\}$ is a synthetic exceedance set drawn from $\mathrm{GPD}(\hat{\xi}_t,\hat{\beta}_t)$, with $|\tilde{\mathcal{Z}}_t|$ matched to $|\mathcal{Z}_t|$.

A discriminator $D_{\psi}(\cdot)$ is trained to distinguish real exceedances $\mathcal{Z}_t$ from synthetic exceedances $\tilde{\mathcal{Z}}_t$. Using the standard logistic GAN objective, the discriminator loss is
\begin{align}
\mathcal{L}_D
&=
-\mathbb{E}_{z\sim\mathcal{Z}_t}\!\big[\log D_{\psi}(z)\big]
-\mathbb{E}_{\tilde{z}\sim\tilde{\mathcal{Z}}_t}\!\big[\log(1-D_{\psi}(\tilde{z}))\big],
\label{eq:disc_loss}
\end{align}
and the generator (parameter MLP) is trained with
\begin{align}
\mathcal{L}_G
&=
-\mathbb{E}_{\tilde{z}\sim\tilde{\mathcal{Z}}_t}\!\big[\log D_{\psi}(\tilde{z})\big],
\label{eq:gen_loss}
\end{align}
which corresponds to a binary cross-entropy (BCE) adversarial objective. Minimizing $\mathcal{L}_G$ updates $f_{\theta}^{(p)}$ through \eqref{eq:generator_comp} so that generated exceedances match the empirical tail distribution.

The online module returns the current tail model $(\hat{u}_t,\hat{\xi}_t,\hat{\beta}_t)$, which can be queried to compute \ac{URLLC}-relevant tail probabilities and quantiles for reliability-aware network decisions.
\section{Numerical Results}
\label{sec:numerical_results}

In this section, we evaluate the proposed \ac{EVT}-guided generative-\ac{AI} framework (\textit{\ac{EVT}-\ac{GAN}}) for channel-tail learning, covering both tail-aware augmentation and online tail-parameter estimation. We compare against widely used generative and non-generative baselines for channel/CSI inference: vanilla \textit{GAN}, \textit{VAE}, \textit{Diffusion}, and \textit{GAN-LSTM} \cite{genai_04}, motivated by prior generative-CSI estimators \cite{genai_csi_01,genai_gan_11,genai_gan_12} and diffusion-based channel modeling/inference \cite{diff_11,diff_12}. We also include non-generative tail estimators: \textit{\ac{MLP}-\ac{KL}}, a feedforward \ac{MLP} trained by minimizing the \ac{KL}-divergence between the implied tail distribution and the empirical tail; \textit{MLP-Bayesian}, an \ac{MLP} trained under a Bayesian \ac{EVT} formulation \cite{bayesian_evt}; and \textit{\ac{MLE}}, a statistical baseline that fits \ac{GPD}/\ac{GEV} parameters from threshold exceedances via maximum likelihood.

\subsection{Experimental Setup and Dataset}
The dataset used in our experiments, described in \cite{evt_02}, was collected inside the engine compartment of a Fiat Linea operating at $60$~GHz under various driving conditions, including a static scenario, driving on a smooth road, and driving on a ramp road. The transmitter and receiver antennas were placed within the engine compartment to capture the effects of engine vibrations on the received power, which are identified as the primary source of extreme events. The antenna locations correspond to realistic placements of the wireless sensors within the compartment. A Vector Network Analyzer (R\&S\textsuperscript{\textregistered} ZVA$67$) was connected to the \ac{Tx} and \ac{Rx} through R\&S ZV-Z196 cables, with a maximum transmission loss of $4.8$~dB. Horn antennas with a $24$~dBi gain and $12^\circ$ vertical beamwidth were used. A total of approximately $10^6$ samples were recorded over a 30-minute period, with a time resolution of $2$~ms. The full dataset was divided into eight stationary batches, each of which was further split into \(8\) sub-stationary batches for training and evaluation purposes.

\subsection{Model Architecture and Training}
The tail-parameter estimator is implemented as a feedforward \ac{MLP} that maps each input window of length $100$ to the corresponding \ac{GPD} parameters $(\hat{\xi}_t,\hat{\beta}_t)$. The network consists of two hidden layers with $64$ neurons each and \textit{ReLU} activations, followed by output transformations using $\tanh$ and \textit{softplus} functions to enforce valid parameter ranges, i.e., $\hat{\xi}_t \in (-0.5,0.5)$ and $\beta>0$. The discriminator is implemented as a three-layer fully connected network with $128$ hidden units and \textit{LeakyReLU} activations, operating on transformed tail samples to distinguish real from generated exceedances. Adversarial training is performed using mini-batches of windowed samples. The generator and discriminator are optimized using the Adam optimizer with learning rates of $2\times10^{-4}$ and $4\times10^{-4}$, respectively.

The threshold estimation network follows a similar \ac{MLP} structure, mapping each input window to a scalar threshold value. It consists of two hidden layers with $64$ neurons and \textit{ReLU} activations, and is trained separately prior to adversarial parameter estimation. The predicted threshold is subsequently used to extract exceedances for both training and evaluation of the tail models. Training is performed using the Adam optimizer with a learning rate of $5\times10^{-5}$, along with a learning rate scheduler and early stopping.



For data augmentation, \ac{EVT}-\ac{GAN} is trained on the full dataset using mini-batches of size $1024$. The generator follows a hybrid design, combining a neural network with two hidden layers of $64$ units (ReLU activations) and a latent input of dimension $16$ for modeling the bulk, together with explicit GPD-based sampling for the tail. The discriminator is implemented as a two-layer fully connected network with $64$ hidden units and LeakyReLU activations. Training is performed for $50$ epochs using the Adam optimizer with a learning rate of \(2\times10^{-4}\) and momentum parameters $(0.5, 0.999)$. 

All models are implemented in \textit{PyTorch} and trained on a \ac{GPU}-enabled environment. The dataset is split into training and evaluation subsets such that both sets contain samples from all stationary regimes.

\subsection{Performance Evaluation}
\begin{figure*}[t]
\centering 

\subfloat[\label{fig:qq-evt-gan}]%
{\includegraphics[width=0.44\columnwidth]{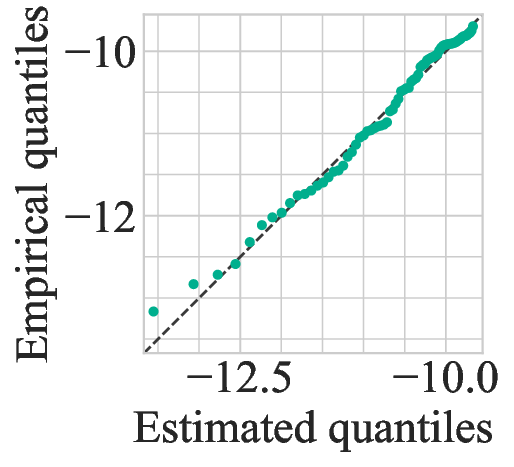}}\hfill
\subfloat[\label{fig:qq-evt-kl}]%
{\includegraphics[width=0.44\columnwidth]{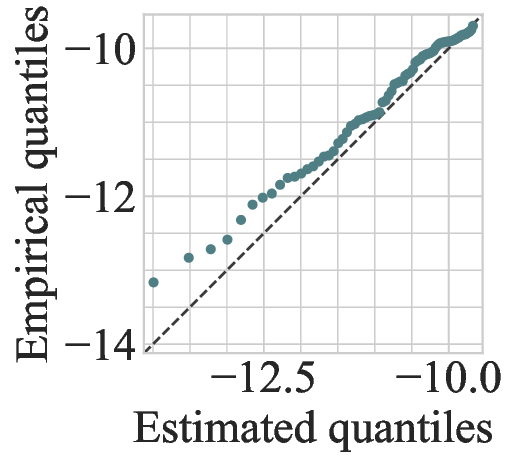}}\hfill
\subfloat[\label{fig:qq-evt-bayes}]%
{\includegraphics[width=0.44\columnwidth]{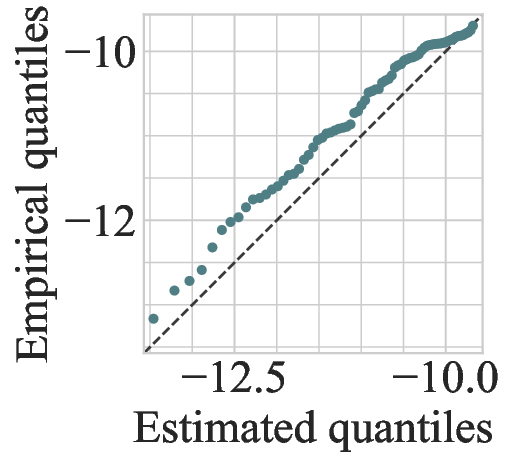}}\hfill
\subfloat[\label{fig:qq-evt-mle}]%
{\includegraphics[width=0.44\columnwidth]{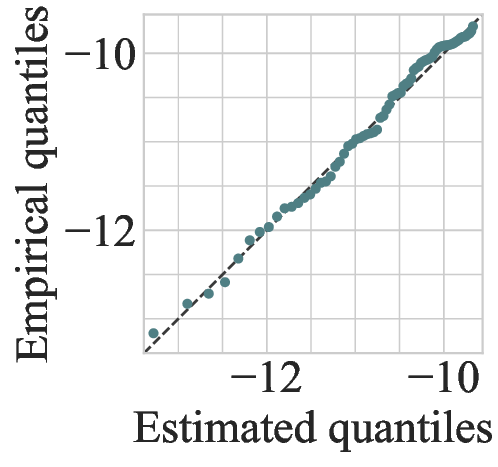}}

\vspace{0.6em}

\subfloat[\label{fig:qq-syn-gan}]%
{\includegraphics[width=0.44\columnwidth]{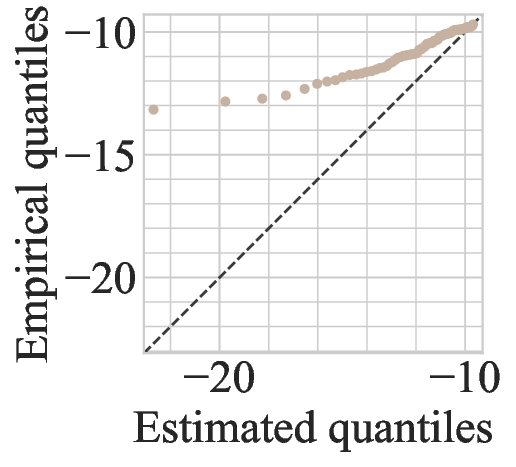}}\hfill
\subfloat[\label{fig:qq-syn-diff}]%
{\includegraphics[width=0.44\columnwidth]{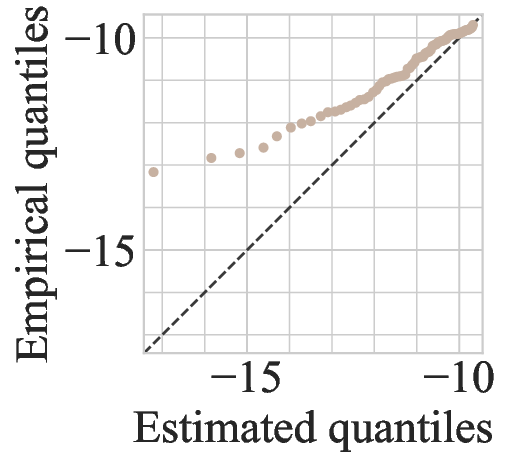}}\hfill
\subfloat[\label{fig:qq-syn-vae}]%
{\includegraphics[width=0.44\columnwidth]{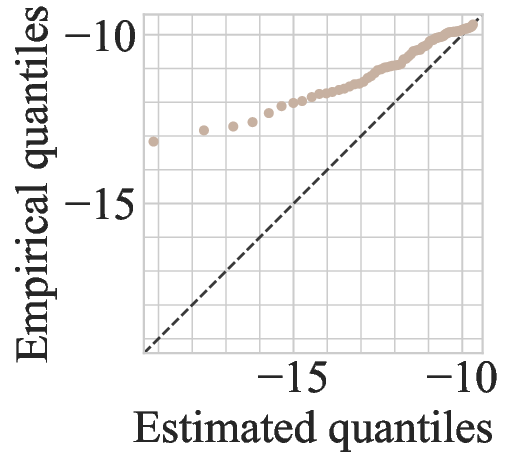}}\hfill
\subfloat[\label{fig:qq-syn-ganlstm}]%
{\includegraphics[width=0.44\columnwidth]{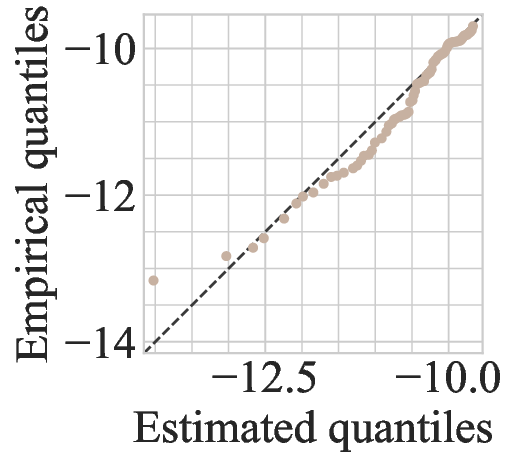}}

\caption{QQ plots comparing empirical vs.\ modeled tail quantiles across EVT-based and generative models: a) EVT-GAN; b) MLP-KL; c) MLP-Bayesian; d) MLE; e) GAN; f) Diffusion; g) VAE; and h) GAN-LSTM.}
\label{fig:qq-all}
\end{figure*}

Fig.~\ref{fig:qq-all} shows the \ac{QQ} plots comparing the estimated tail distributions with the empirical tails. The proposed \ac{EVT}-\ac{GAN} parameter estimator (Fig.~\ref{fig:qq-evt-gan}) achieves better alignment with the empirical distribution compared to the KL-based (Fig.~\ref{fig:qq-evt-kl}) and Bayesian \ac{EVT} estimators (Fig.~\ref{fig:qq-evt-bayes}), indicating more accurate modeling of the tail behavior. The estimation quality of \ac{EVT}-\ac{GAN} is also comparable to, and in some cases exceeds, that of the classical \ac{MLE} approach (Fig.~\ref{fig:qq-evt-mle}), while requiring significantly fewer samples. In particular, \ac{MLE} requires approximately $120\times$ more samples to achieve a similar level of accuracy in tail estimation. The first row (Fig.~\ref{fig:qq-evt-gan}-Fig.~\ref{fig:qq-evt-mle}) corresponds to EVT-based models, while the second row (Fig.~\ref{fig:qq-syn-gan}-Fig.~\ref{fig:qq-syn-ganlstm}) presents generative models. The results show that generative approaches, including Vanilla \ac{GAN} (Fig.~\ref{fig:qq-syn-gan}), diffusion models (Fig.~\ref{fig:qq-syn-diff}), \ac{VAE} (Fig.~\ref{fig:qq-syn-vae}), and \ac{GAN}-\ac{LSTM} (Fig.~\ref{fig:qq-syn-ganlstm}), struggle to accurately capture the behavior of extreme quantiles. Among these models, \ac{GAN}-\ac{LSTM} performs relatively better, which can be attributed to its ability to capture temporal dependencies in the data, leading to improved modeling of extreme events compared to purely feedforward generative models. 

\begin{figure}[t]
\centering
\subfloat[\label{fig:box-ks}]%
{\includegraphics[width=0.33\textwidth]{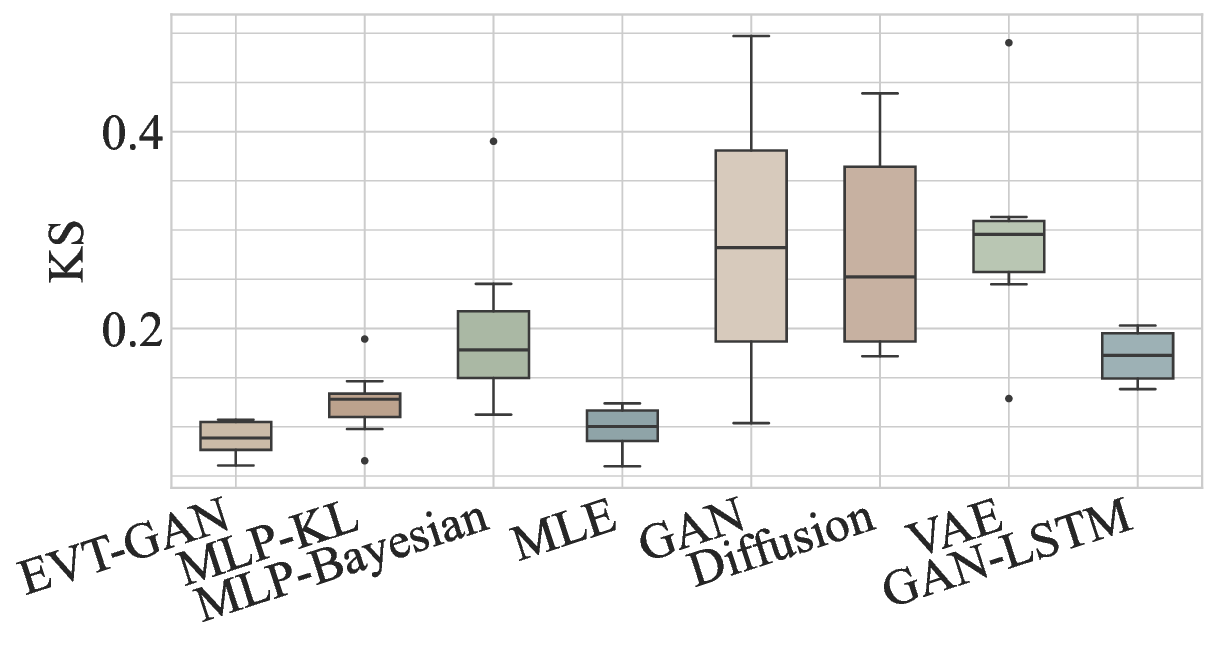}}\\
\subfloat[\label{fig:box-mae}]%
{\includegraphics[width=0.33\textwidth]{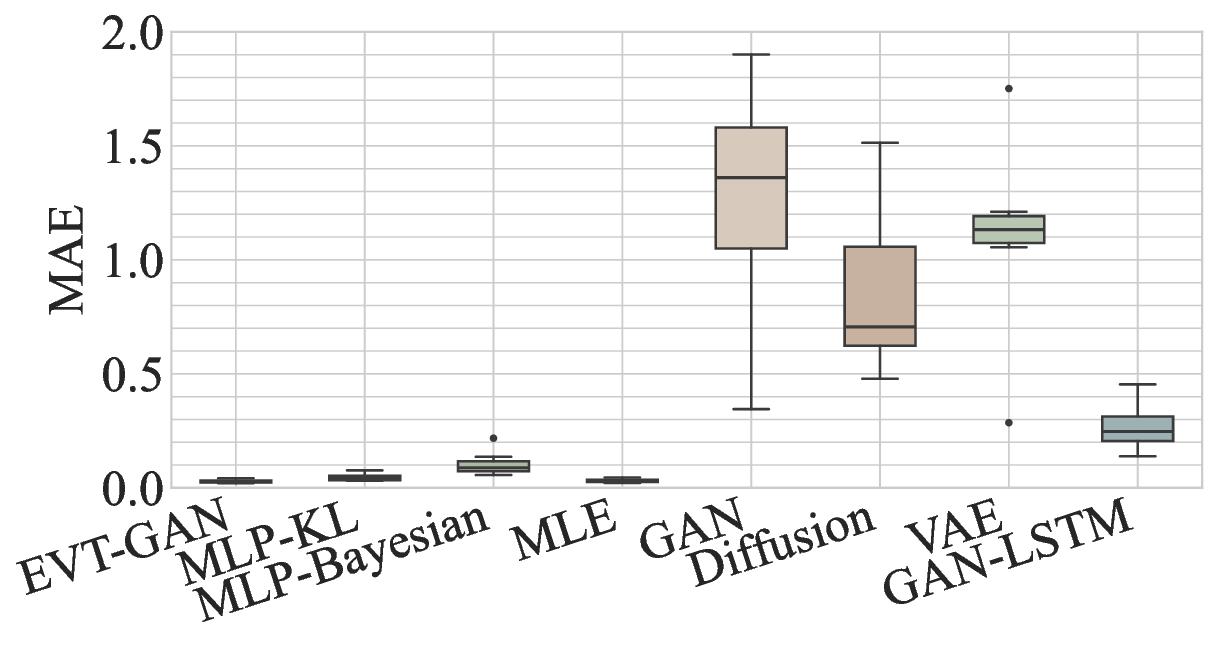}}
\caption{Metric distributions across all models: a) KS; and b) MAE.}
\label{fig:metric-box-ks-mae}
\end{figure}

Fig.~\ref{fig:metric-box-ks-mae} presents the distribution of performance metrics across all test samples and stationary regimes. The proposed \ac{EVT}-\ac{GAN} achieves consistently lower \ac{KS} values (Fig.~\ref{fig:box-ks}) compared to the \ac{KL}-based and Bayesian estimators, indicating a closer match between the estimated and empirical tail distributions. Similarly, in terms of \ac{MAE} (Fig.~\ref{fig:box-mae}), \ac{EVT}-\ac{GAN} demonstrates reduced estimation error and improved stability across different regimes. The results also show that \ac{EVT}-based methods exhibit significantly tighter distributions compared to generative models, reflecting more reliable and consistent tail estimation. In contrast, generative approaches, including \ac{GAN}, diffusion models, \ac{VAE}, and \ac{GAN}-\ac{LSTM}, display larger variability and higher errors, particularly in challenging regimes. Overall, \ac{GAN}-\ac{LSTM} shows relatively improved performance due to its ability to capture temporal dependencies, though it still falls short of \ac{EVT}-based approaches. 

\begin{table}[t]
\centering
\footnotesize
\caption{Tail-modeling comparison across different models.}
\label{tab:tail-model-comparison-common}
\setlength{\tabcolsep}{4pt}
\renewcommand{\arraystretch}{1.15}

\begin{tabular}{
l|
S[table-format=5.0]
S[table-format=1.2]
S[table-format=1.2]
S[table-format=1.2]
S[table-format=1.2]
S[table-format=1.2]
}
\toprule
{Model} & {N $\downarrow$} & {KS $\downarrow$} & {MSE $\downarrow$} & {RMSE $\downarrow$} & {MAE $\downarrow$} & {PPCC $\uparrow$} \\
\midrule
EVT-GAN          & \textbf{ 72}    & \textbf{ 0.077} & \textbf{ 0.001} & \textbf{ 0.034} & \textbf{ 0.026} & \textbf{ 0.994} \\
MLP-KL           & \textbf{ 72}    & 0.130 & 0.005 & 0.071 & 0.063 & 0.992 \\
MLP-Bayesian     & \textbf{ 72}    & 0.176 & 0.012 & 0.110 & 0.099 & 0.985 \\
MLE          & \textbf{ 72}    & 0.091 & \textbf{ 0.001} & 0.040 & 0.029 & 0.993 \\
Diffusion     & 6993  & 0.181 & 0.946 & 0.973 & 0.660 & 0.984 \\
GAN     & 15547 & 0.317 & 4.878 & 2.208 & 1.474 & 0.967 \\
VAE     & 15501 & 0.295 & 2.828 & 1.681 & 1.211 & 0.987 \\
GAN-LSTM & 909   & 0.138 & 0.039 & 0.197 & 0.138 & 0.981 \\
\bottomrule
\end{tabular}
\end{table}

Table \ref{tab:tail-model-comparison-common} summarizes the quantitative performance of all models across common evaluation metrics, including \ac{KS}, \ac{MSE}, \ac{RMSE}, \ac{MAE}, and \ac{PPCC}. The proposed \ac{EVT}-\ac{GAN} achieves the best overall performance, with the lowest \ac{KS} and \ac{MAE} values and the highest \ac{PPCC}, indicating superior agreement with the empirical tail distribution. Compared to other EVT-based approaches, \ac{EVT}-\ac{GAN} consistently improves estimation accuracy, while maintaining stable performance across all metrics. The column $N$ denotes the number of tail samples below the threshold used for evaluation. \ac{EVT}-based methods operate with a limited number of true exceedances ($N=72$), reflecting realistic tail observations. In contrast, generative models produce a significantly larger number of samples in the tail region; however, these samples are not accurately distributed and lack proper balance, leading to degraded performance across all metrics. This is reflected in the higher error values and lower goodness-of-fit measures observed for GAN, \ac{VAE}, diffusion, and \ac{GAN}-\ac{LSTM} models.

\begin{figure}[t]
\centering

\subfloat[\label{fig:bars-params}]%
{\includegraphics[width=0.33\textwidth]{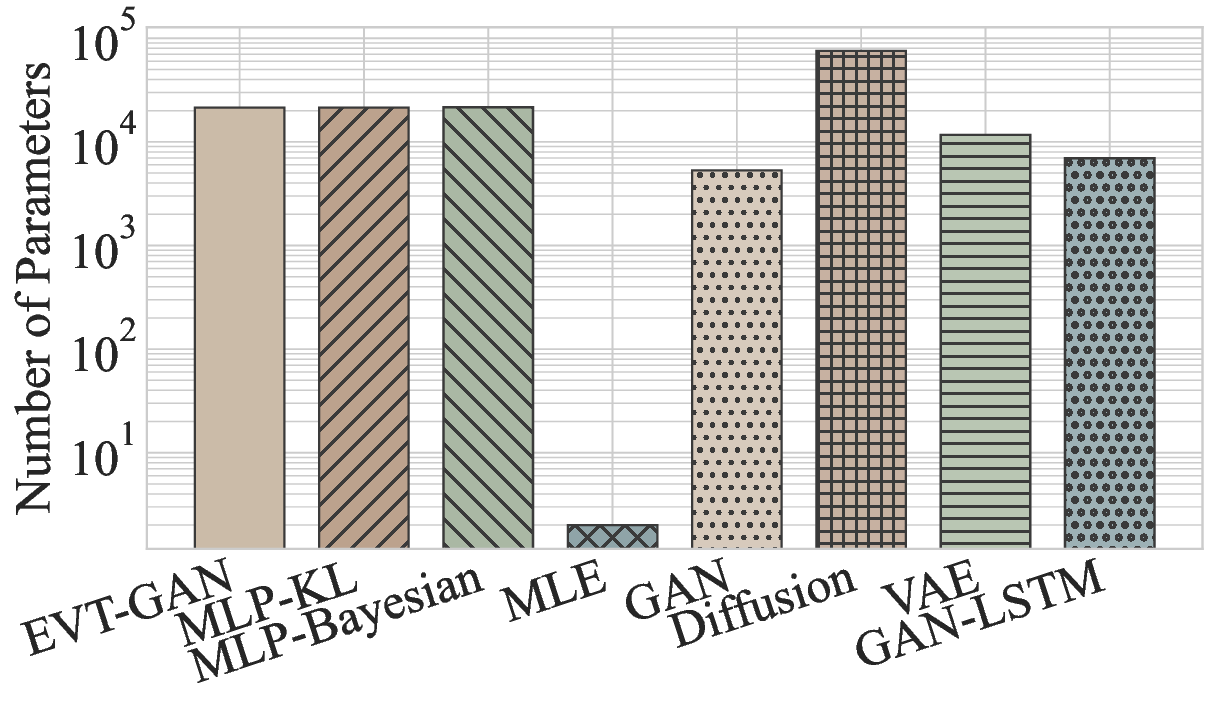}}\\
\subfloat[\label{fig:bars-infer}]%
{\includegraphics[width=0.33\textwidth]{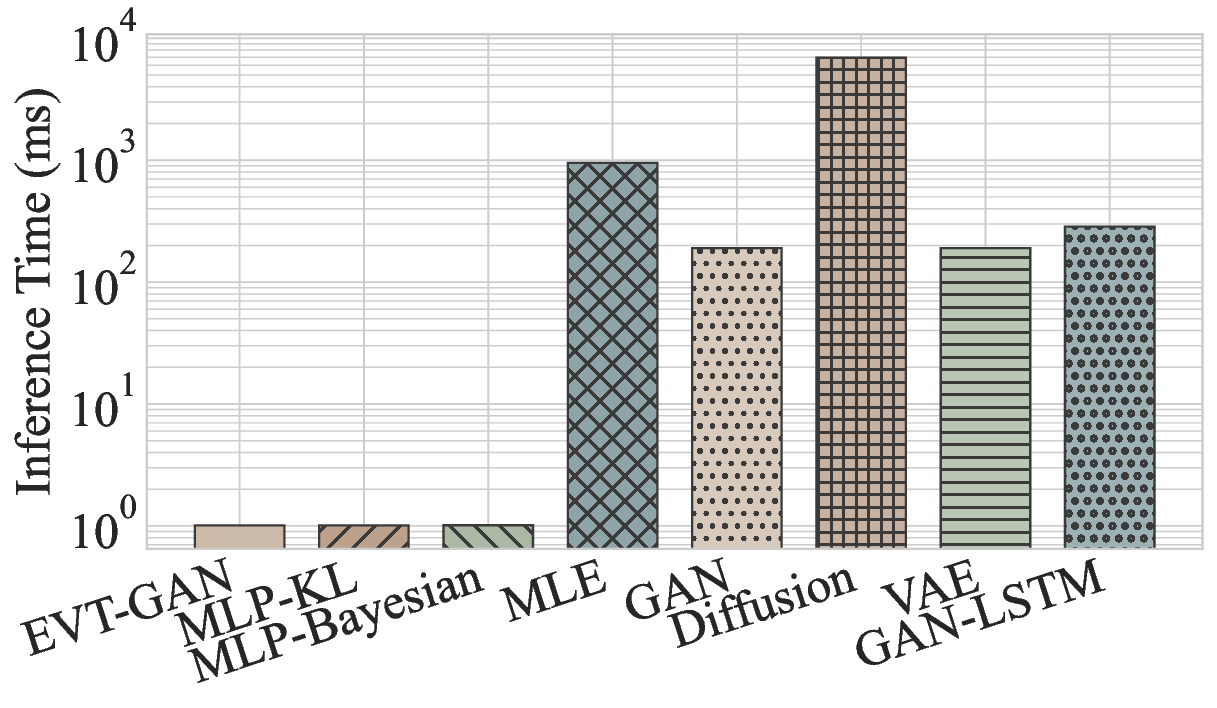}}

\caption{Computational cost comparison across tail-estimation models: a) Parameter count; and b) Inference time.}
\label{fig:bars-params-infer}
\end{figure}

Fig.~\ref{fig:bars-params-infer} compares the computational cost of all models in terms of parameter count (Fig.~\ref{fig:bars-params}) and inference time (Fig.~\ref{fig:bars-infer}), both shown on a logarithmic scale. The proposed \ac{EVT}-\ac{GAN}, along with the \ac{KL}-based and Bayesian estimators, exhibits similar model sizes and inference times, as all three approaches rely on the same underlying network architecture and operate on input windows of fixed size to jointly estimate threshold and tail parameters. The differences among these methods arise primarily from their training objectives rather than their computational complexity. In contrast, the \ac{MLE}-based approach has a negligible parameter count due to its purely statistical nature; however, it requires collecting a large number of samples ($\approx$ $9.5\times10^{4}$) to obtain sufficient tail observations for reliable estimation, which significantly increases its effective latency. Generative models, including \ac{GAN}, \ac{VAE}, diffusion, and \ac{GAN}-\ac{LSTM}, also incur higher computational costs, as they must generate a large number of samples to ensure adequate coverage of the tail region. This additional sampling requirement leads to increased inference time without guaranteeing accurate tail estimation.

\begin{figure}[t]
\centering

\subfloat[\label{fig:tc-with-thr}]%
{\includegraphics[width=0.23\textwidth]{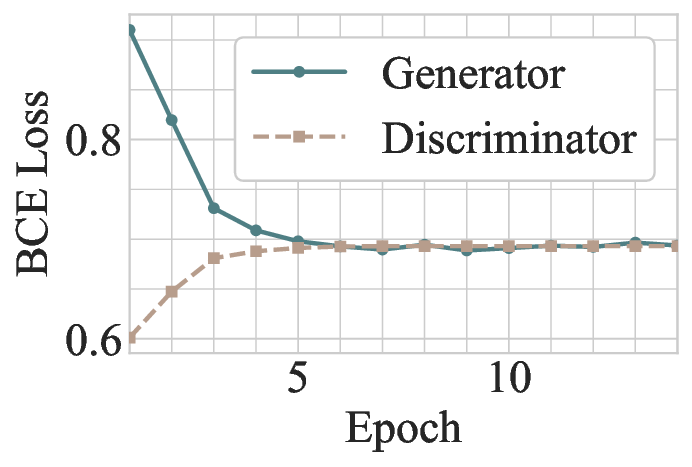}}\hfill
\subfloat[\label{fig:tc-no-thr}]%
{\includegraphics[width=0.23\textwidth]{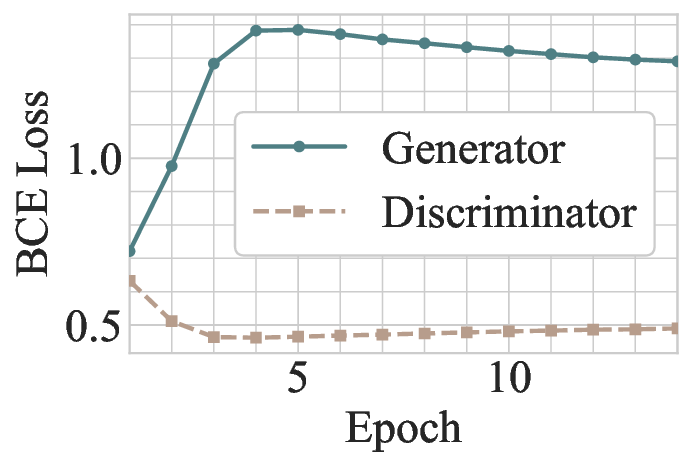}}
\caption{Training curves (BCE loss) for the generator and discriminator: a) EVT-GAN with Threshold MLP; and b) EVT-GAN without Threshold MLP.}
\label{fig:training-curves}
\end{figure}

Fig.~\ref{fig:training-curves} illustrates the training dynamics of the proposed \ac{EVT}-\ac{GAN}, comparing the cases with and without the threshold estimation network. When the threshold \ac{MLP} is included (Fig.~\ref{fig:tc-with-thr}), both generator and discriminator losses exhibit stable convergence behavior, indicating effective adversarial training and consistent tail modeling. In contrast, when a constant threshold is used across all regimes (Fig.~\ref{fig:tc-no-thr}), the training process becomes less stable, with increased fluctuations in the loss curves. This behavior highlights the importance of the threshold estimation network, as the optimal threshold varies significantly across different stationary regimes. Without adapting the threshold to each input window, the extracted tail samples become inconsistent, leading to degraded training dynamics and less reliable parameter estimation. These results confirm that accurate threshold modeling is a critical component of the proposed framework.

\begin{figure}
    \centering
    \includegraphics[width=0.65\linewidth]{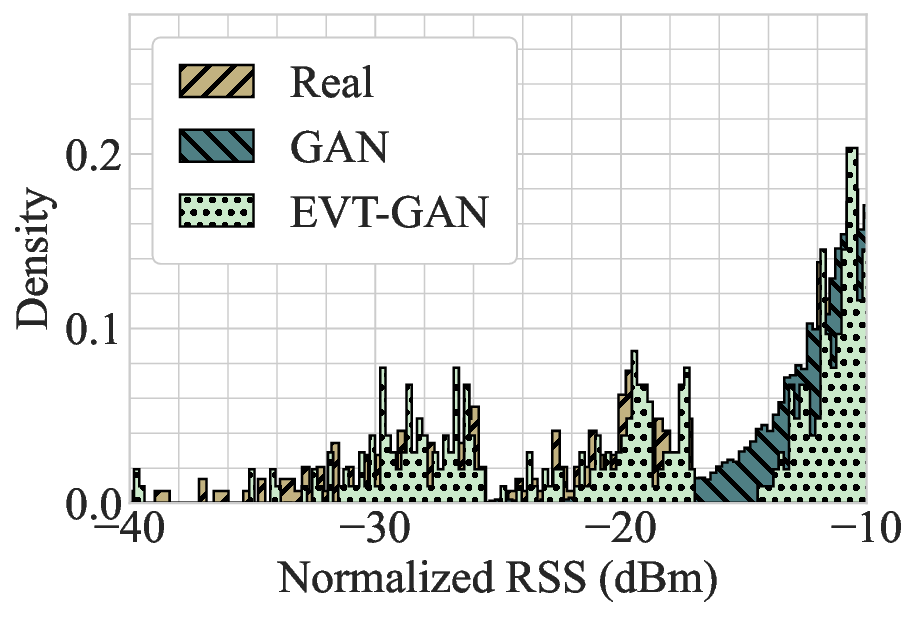}
    \caption{The distribution of the tail samples generated by EVT-GAN and GAN; RSS refers to received signal strength.}
    \label{fig:data_aug}
\end{figure}

Fig.~\ref{fig:data_aug} compares the tail distributions generated by the proposed \ac{EVT}-\ac{GAN} and a standard Vanilla GAN for data augmentation. The results show that the Vanilla \ac{GAN} fails to adequately cover the extreme tail region, with generated samples concentrated around higher-density areas and lacking representation of rare events. In contrast, the \ac{EVT}-\ac{GAN} accurately reproduces the tail distribution by explicitly modeling extreme values through \ac{GPD}-based sampling. This difference is critical, as the primary objective of the proposed framework is accurate tail modeling. The inability of conventional generative models to capture deep tails leads to insufficient and biased training samples for tail estimation. By generating samples that closely follow the true tail behavior, \ac{EVT}-\ac{GAN} provides reliable augmentation that directly supports the learning of tail parameters.

It is worth noting that, although the MLE-based approach has a negligible parameter count, it is not suitable for real-time operation, as it must accumulate a large number of exceedances to produce reliable estimates, resulting in inference latency that is several orders of magnitude higher than the proposed approach. In contrast, EVT-GAN performs direct, window-based parameter estimation, enabling real-time inference while remaining fully trainable, supporting online adaptation and improved generalization across varying regimes. The proposed model is also lightweight and GPU-efficient, sharing a compact architecture with other MLP-based estimators, while outperforming more complex generative models such as GANs, VAEs, diffusion models, and GAN-LSTMs, which incur higher computational costs without improving tail estimation. 


\section{Conclusions}
\label{sec:conclusions}

In this paper, we have proposed a novel approach for designing URLLC systems by integrating generative AI with EVT. We demonstrate that incorporating generative AI into EVT-based statistical tools, including optimal threshold selection, tail statistics modeling, and validation, significantly accelerates the parameter estimation process. Moreover, the EVT-based generative AI approach enables effective data augmentation, enriching extreme quantiles and improving overall accuracy. Using an experimental dataset, we show that combining EVT with generative AI improves the estimation of extreme quantiles and achieves \(0.994\) accuracy, based on PPCC criteria, in online estimation of channel tail distribution parameters. In the future, we aim to extend the proposed framework to establish theoretical guarantees on the reliability and to validate the generalizability of the algorithms in other scenarios, including unmanned aerial vehicle (UAV) measurements.

\balance 

\ifCLASSOPTIONcaptionsoff
  \newpage
\fi
\bibliographystyle{ieeetr}
\bibliography{GenAI_EVT_ICT_Conf}

\end{document}